\documentclass[aps,preprintnumbers,amsmath,amssymb,floatfix,groupedaddress,nofootinbib]{revtex4}

\usepackage{enumerate}
\usepackage{graphicx}
\usepackage{color}
\usepackage[utf8]{inputenc}
\usepackage{enumitem}
\usepackage{appendix}
\usepackage{ulem}

\usepackage{fontenc}  
\usepackage{amsmath}

\newcommand{\beq}{\begin{equation}}
\newcommand{\eeq}{\end{equation}}
\newcommand{\bea}{\begin{eqnarray}}
\newcommand{\eea}{\end{eqnarray}}
\newcommand{\ba}{\begin{array}}
\newcommand{\ea}{\end{array}}

\newcommand{\bef}{\begin{figure}}
\newcommand{\eef}{\end{figure}}

\begin{document}

\title{Comment on ``There is No Quantum World" by Jeffrey Bub. }

\author{Philippe Grangier}
\affiliation{Laboratoire Charles Fabry, IOGS, CNRS, 
Universit\'e Paris Saclay, F91127 Palaiseau, France.}

\begin{abstract}

In a recent preprint \cite{bub} Jeffrey Bub presents a discussion 
of neo-Bohrian interpretations of quantum mechanics, and also of von Neumann’s work on infinite tensor products \cite{JvN39}. He rightfully writes that this work provides a theoretical framework that deflates the measurement problem and justifies Bohr’s insistence on the primacy of classical concepts. But then he rejects these ideas, on the basis that the infinity limit is ``never reached for any real system composed of a finite number of elementary systems". In this note  we 
present opposite views on 
two major points: first, admitting mathematical infinities in a physical theory is not a problem, if properly done; second, the critics of \cite{MP1,MP2,MP3} comes with a major misunderstanding of these papers: they don’t ask about ``the significance of the transition from classical to quantum mechanics", but they start from a physical ontology where classical and quantum physics need each other from the beginning. This is because they postulate that a microscopic physical object (or degree of freedom) always appears as a quantum system, within a classical context. Here we argue why this (neo-Bohrian) position makes sense.  

\end{abstract}

\maketitle

\section{Introduction: disagreements and proposed clarifications. }

Bub's paper \cite{bub} raises important issues about the interpretation of quantum mechanics and about the status of classical concepts. We agree with much of his historical and philosophical discussion, and share his interest in von Neumann's infinite‑product framework. However we disagree on two major issues.
\\

The first disagreement is 
about the methodological statement that the infinite limit would be irrelevant
to physics because no real system attains it exactly. 
Despite this, it is standard practice in physics to use mathematical limits as idealizations that reveal robust,
qualitative behavior of large but finite systems. Well-known examples include the thermodynamic limit in statistical mechanics, which explains phase transitions and  emergent order,
and also continuum limits in mechanics and field theory, which allow differential equations to  model media composed of discrete atoms.
Therefore, arguing that an infinite construction is irrelevant because it is never literally attained conflates ontological literalism with methodological usefulness. The infinite tensor product can be used in the same spirit: it provides a mathematically controlled idealization that describes the {\bf structural} features (super-selection sectors, stable records) that macroscopic systems exhibit. More details will be given below. 
\\

The second disagreement stems from quite differing ontological emphases. In our CSM (Contexts, Systems and Modalities) approach \cite{MP1,MP2,MP3}
the basic ontology is not ``a
purely quantum world'' or ``a purely classical world'' but \emph{quantum systems} that are described
within \emph{classical contexts}. This is neither a retreat to naive classical realism, nor a mere revival of Bohr's complementarity: 
it is a realist stance that recognizes the irreducible role of contexts (macroscopic apparatus,
classical records) in the operational meaning of quantum states and probabilities. From this starting
point, the mathematical framework that includes infinite tensor products and the associated sector
structure becomes a natural tool to model how macroscopic contexts stabilize definite outcomes.
Then quantum and classical physics can be described in a unified algebraic formalism - without one ``emerging" from the other. 
\\

In order to give some flesh to the statements above, we quote in italics some paragraphs extracted from Bub's paper \cite{bub}, and provide specific comments and references for each one. 

\section{Commented quotes from Bub's preprint}

\subsection{A quote from von Neumann \cite{JvN}}
\vspace{-2 mm}
{\it Another way to put it is that if you take the case of an orthogonal space,
those mappings of this space on itself, which leave orthogonality intact,
leave all angles intact, in other words, in those systems which can be used
as models of the logical background for quantum theory, it is true that as
soon as all the ordinary concepts of logic are fixed under some isomorphic
transformation, all of probability theory is already fixed.}
\\

Comment: This enigmatic sentence by John von Neumann was visionary, given that he told this in 1954, before knowing Uhlhorn’s theorem (about leaving orthogonality intact) and Gleason’s theorems (about an isomorphic transformation fixing all of (quantum) probability theory). These issues are discussed in details in \cite{Gleason, Uhlhorn}. 
\vspace{-2 mm}
\subsection{The sort of realist explanation we are familiar with - or not}
\vspace{-2 mm}
{\it On this view, quantum mechanics does not provide a representational explanation
of events. Noncommutativity or non-Booleanity makes quantum mechanics quite unlike
any theory we have dealt with before, and there is no reason, apart from tradition, to
assume that the theory should provide the sort of realist explanation we are familiar
with in a theory that is commutative or Boolean at the fundamental level.}
\\

Comment: Yes! However, giving up ``the sort of realist explanation we are familiar with" does not mean to give up any kind of realist explanation.  Contextual objectivity \cite{CO2002} provides a framework which is fully eligible as a realist explanation, though not as the classical one we are familiar with. It thus crucial not to confuse physical realism with classical, reductionist realism
in order to catch what quantum physics is telling us. . 
\vspace{-2 mm}
\subsection{The interaction has something to do with us - or not}
\vspace{-2 mm}
{\it For a physical system a measurement is a dynamical interaction with a second
system that we characterize as a measuring instrument. The fact that we extract 
information about the measured system from the interaction has something to do with us,
not the measured system. There should be no change in the nature of the dynamical
behavior of the measured system just because we choose to regard the interaction as a
measurement, so a special dynamics for measurement is unacceptable.}
\\

Comment: No! Why claiming that ``the interaction has something to do with us" ? Obviously there is no physics without physicists, but in our view there is a physical world that does not care about physicists: it  was there before and is likely to be there after there is any physicist to look at it. Even now, quantum measurements can be performed by automated devices, that will operate the same way, even very far from any physicist. What remains true is that the physical world comprises microscopic  systems, and macroscopic contexts, that are incommensurable by many orders of magnitude. This is why ``a special dynamics for measurement" is required \cite{CSM1,CSM2}. The physics done by physicists need to care about this situation, even though the physical world itself does not care. 
\vspace{-2 mm}
\subsection{When is ‘so large’ large enough?}
\vspace{-2 mm}
{\it Van Den Bossche and Grangier are clear about this: ‘In principle the radical change of behavior occurs in the $N \rightarrow \infty$ limit only...’ But they muddy the issue by saying that ‘even before reaching the $N \rightarrow \infty$ limit, the sectorisation behavior sets in and converts the pure state in an effective mixed state,’ and ‘at some point, the number of electrons fed into the avalanche is so large that it results in a macroscopic change that cannot be ignored.’  The crucial words here are ‘effective’ and ‘at some point’ and ‘so large.’ What does ‘effective’ mean here, and at what point is ‘so large’ large enough?}
\\

Comment: In such discussions it is crucial to tell what are the starting and ending points. In our approach the existence of physical objects which are quantum systems within classical contexts is postulated at the beginning, with specific properties \cite{CO2002,CSM1,CSM2},  and the purpose of the mathematical formalism is to describe this situation, not to prove any kind of  ``emergence". Then the standard (type I) quantum formalism describes the behavior of the systems, whereas the  (sectorized, typically type III) infinite limit matches the behavior of contexts. Now, where is the ``cut" between the two ? The argument above just tells that it has to be somewhere, and where it is actually cannot be told in an absolute way, but requires to specify a measurement scheme with interactions  involving a system and a context, initially with a size $N$. Then it can be shown using algebraic techniques *applied to the specific measurement scheme* that the predicted behavior when $N \rightarrow \infty$ is exponentially close to the  behavior predicted for infinite $N$. 
This is a satisfactory situation from our point of view, physically and mathematically, much better anyway than any kind of cramped extrapolation of the type I formalism to the entire universe. 
\vspace{-2 mm}
\subsection{Getting definite probabilities within a context}
\vspace{-2 mm}
{\it Negligible interference is not zero interference. There is always some measurement
that could, in principle, reveal an interference effect, which is to say that the Hilbert
space of a composite system composed of N constituent elementary systems does not
sectorize for any finite N and von Neumann’s problem remains. There is still no explanation 
about how something that is indeterminate or indefinite can become definite
in a measurement process, how something that is neither true nor false in the quantum
state of the measured system can become true or false in a measurement, so that the
Born probabilities that initially refer to indefiniteness can be understood as probabilities 
about our ignorance of what is actually the case.}
\\

Comment: It is not true that ``there is still no explanation": There is an explanation, but it relies on an ontological postulate, telling that physical objects, as we can study them, are quantum systems within classical contexts. There are many reasons, empirical, logical, and mathematical, to adopt this point of view, and as long as it is rejected, physics and physicists are trapped in a cramped conceptual framework. On the other hand, when admitting a description based on systems within contexts, 
one can reconstruct standard quantum mechanics in an inductive way, as done in 
\cite{Gleason, Uhlhorn}. Then (and only then), using the extended mathematical framework introduced in by von Neumann in \cite{JvN39}, one can get an unified picture of classical and quantum physics \cite{MP1,MP2,MP3}. 
\vspace{-2 mm}
\subsection{A major issue: using mathematical infinites in physics}
\vspace{-2 mm}
{\it There is, however, a solution outside quantum mechanics at the $N \rightarrow \infty$ limit, where
the properties of macrosystems modeled as composite physical systems depend on the
collective behavior of their elementary constituents and are insensitive to adding or 
removing any finite number of elementary systems. But since the limit is never reached
for any real system composed of a finite number of elementary systems, there is no
number N such that the partial Boolean algebra of a composite quantum system with
more than N constituent elementary systems becomes Boolean. So the Boolean macro-
world is not reducible to the non-Boolean quantum level in the usual sense.}
\\

Comment: This statement makes a  confusion about the relationship between physics 
and mathematics. The mathematical formalism, in the way the physicists use it, 
is a language able to describe nature - like an improved version of standard 
language. But describing is not being isomorphic: there are
many examples where mathematical infinities are relevant for physics, e.g. just for properly 
calculating derivatives and integrals, and also in thermodynamics for instance. Nobody 
claims that infinitesimals ``really exist", or that a system undergoing a phase transition 
``really contains" an infinite number of particles. But using these infinite limits is quite standard,
and there is no serious reasons for not doing the same with quantum mechanics. 
It is  noteworthy  that what is changing in the quantum infinite  limit is not only a calculated number,  but a whole algebraic structure. Nevertheless, since Cantor there is nothing here to frighten a mathematician, and so there should be nothing either to frighten a physicist \cite{MP1,MP2,MP3}. As an illustration some basic ideas from operator algebras are presented in the Appendix. 
\vspace{-2 mm}
\section{Concluding remarks}
\vspace{-2 mm}
Bub's paper is a valuable contribution to the foundations literature and rightly draws attention
to von Neumann's deep work on infinite tensor products. The present reply is not a rejection of his concerns but a plea for a different methodological stance: mathematical idealizations such as infinite tensor products are
legitimate and often indispensable tools in physics. When combined with an ontology that treats
systems and contexts as co‑primary, the infinite‑product framework helps to explain why macroscopic
definiteness and classical descriptions are stable and operationally meaningful, even though every
actual system is finite.

The CSM approach to quantum mechanics that we introduced in \cite{CO2002,CSM1,CSM2}, completed by \cite{Gleason,Uhlhorn} and \cite{MP1,MP2,MP3} can certainly be considered as neo-Bohrian, or neo-Copenhagian, in the sense that  it is closer to Bohr’s ideas than to any other 
later interpretation like Everett’s or Bohm’s.  However, looking more closely, it is also quite different from Bohr: this is not a big surprise, since 100 years later we know with certainly many theoretical and experimental facts that Bohr could only guess. Among other differences, we never refer to Bohr’s complementarity, which was in our opinion loosely defined. On the other hand, we give essential importance to the combination of quantization and contextuality: actually, contextual quantization is in our approach the crucial feature of quantum physics, from which we can induce the quantum formalism, and remove ``paradoxes" mixing up micro and macro \cite{entropy,completing}. 

So, is there a quantum world ? This question has little meaning, but objects made of quantum systems within classical contexts belong for sure to the physical world. 
These ontological ideas, associated with the general formalism of operator algebras going beyond 
textbook quantum mechanics, allow us to set up a combined description of systems and contexts.  This provides in our opinion several interesting avenues to be explored further. 
\vspace{- 2mm}
\subsection*{Acknowledgements} 
\vspace{-2mm}
The author thanks colleagues for helpful comments. Some input from Copilot has been used for edition. 


\appendix 
\section{Von Neumann's infinite tensor product for dummies}
\label{app:vn_accessible}
\vspace{- 2mm}
This appendix gives a simple 
account of the infinite tensor‑product construction
and of the \emph{tail algebra} (also called asymptotic algebra) that plays a central role in the
sectorization picture. The aim is to convey the ideas and physical intuition without assuming the reader
already knows the full operator‑algebra machinery. For the original
ideas see von Neumann \cite{JvN39}, and for a more recent treatment see e.g. \cite{KR}. We exploit physically these ideas in \cite{MP1,MP2,MP3}. 

\subsection{Why consider an infinite tensor product}
\vspace{-2 mm}
Physicists often use infinite limits as \emph{idealizations} that reveal robust behavior of very large but
finite systems. The infinite tensor product is such an idealization tailored to composite quantum systems:
it isolates algebraic structures that become effectively stable when the number of constituents is very large.
Two motivating points are: (i)  \textbf{Stability of macroscopic records.} Macroscopic measurement outcomes are stable under small,
  local perturbations; the infinite limit makes this stability manifest as a structural feature (sectors).
(ii) \textbf{Separation of scales.} Local observables (probing a finite region) and global, asymptotic observables
  (probing the far tail) behave differently; the infinite construction separates these roles cleanly.
\vspace{-2 mm}
\subsection{A short sketch of the construction}
\vspace{-2 mm}
Let us start from a finite composite sequence of many  systems. Concretely, let
\(\mathcal H^{(n)}=\bigotimes_{k=1}^n \mathcal H_k\) be the Hilbert space of the first \(n\) subsystems,
with each factor \(\mathcal H_k\) finite dimensional for simplicity. We choose a reference product state
(e.g., a fixed unit vector in each factor) and embed \(\mathcal H^{(n)}\) into \(\mathcal H^{(n+1)}\)
by tensoring with the reference vector on the \((n+1)\)-th factor. The algebraic union of the finite
operator algebras (operators that act nontrivially only on finitely many factors) is then represented on
the Hilbert space obtained from the reference state; taking the weak operator closure yields a von Neumann
algebra that encodes the infinite system. 

This is the route von Neumann followed in \cite{JvN39}.
More recently, many functional‑analysis tools have been elaborated, such as the Gelfand-Naimark-Segal (GNS) construction \cite{KR}.
The important
point  is that the infinite construction is a well‑defined limit of finite systems that preserves the physically relevant algebraic relations.
\vspace{-2 mm}
\subsection{What the tail algebra is, and how sectors arise from it.}
\vspace{-2 mm}
Intuitively, the \textbf{tail algebra} consists of observables that act trivially on any fixed finite block
but may act nontrivially on arbitrarily far parts of the system. Think of an infinite spin chain: a local
spin operator acts on site 1 or site 2, etc.; a tail observable is insensitive to any finite initial segment and
only probes the asymptotic remainder.
More concretely, a \textbf{local observable} affects only a finite set of factors (for example, the first \(m\) spins).
A \textbf{tail observable} commutes with every local observable in the sense that, for any fixed local
  region, there exist tail observables supported entirely outside that region.
The tail algebra is generated by such tail observables; it captures the global, asymptotic degrees of freedom.

The tail algebra can have a nontrivial center, made of commuting elements that behave like classical labels.
Central projections in the tail algebra split the representation into orthogonal components, called
\textbf{sectors}. 

Each sector corresponds to a macroscopically distinct configuration of the far tail
(for example, different values of a macroscopic order parameter). Within a given sector, local measurements
have well‑defined statistics that are stable under local perturbations.
Physically, the
sectors play the role of labelling \emph{macroscopic contexts}: once the system is in a sector, local operations  cannot easily move it to another sector.
For very large finite systems the same decomposition is only approximate, but interference between
  different approximate sectors is suppressed to negligible levels.
  This makes the infinite limit meaningful, both mathematically and physically. 
\vspace{-2 mm}
\subsection{Concrete example: an infinite qubit chain}
\vspace{-2 mm}
Consider an infinite chain of qubits with a reference product state \(|0\rangle^{\otimes\infty}\).
Local observables act on finitely many sites. Tail observables include operators that measure the
asymptotic frequency of \(|1\rangle\) occurrences far down the chain (made precise by limits of averages).
Different limiting frequencies label different sectors: one sector might correspond to ``asymptotic density
of ones equals 0'', another to ``density equals 1/2'', and so on. In the infinite idealization these labels
are sharp; for large finite chains the frequencies concentrate sharply around their expected values, so the
same labels are effectively stable.

This simple picture shows why the tail algebra and sectors are not exotic: they formalize the intuitive idea
that macroscopic averages (or other asymptotic observables) behave classically and can serve as stable records.
\vspace{-2 mm}
\subsection{From finite systems to infinite idealizations}
\vspace{-2 mm}
Two features connect the idealization to experiments:

(i)  \textbf{Approximate sectorization.} For any physically realistic finite \(N\) one can identify observables
  that behave as if sectors existed: interference terms between macroscopically distinct outcomes are
  exponentially small in \(N\) in concrete models. Thus the infinite limit explains why finite systems behave
  as if they had superselection sectors.
  
(ii)  \textbf{Robustness under perturbations.} The tail algebra construction makes explicit why macroscopic
  records are robust: local perturbations cannot change the tail labels, so records persist on experimentally
  relevant timescales. Overall, sectorization brings ``infinite systems" back to classical contexts. 

\subsection{Why this helps with the measurement discussion}
\vspace{-2 mm}
The sector picture supplies a structural reason for the apparent ``collapse"  during a measurement:
the global state of system plus apparatus can be decomposed into components
labelled by tail projections (sectors), and local observers interacting with the apparatus effectively
find themselves in one such component. 
The infinite idealization makes the decomposition exact; for
large finite apparatuses the decomposition is an excellent approximation. This is analogous to how the
thermodynamic limit explains phase stability even though real systems are finite.
\vspace{2 mm}

For a  follow up, the original construction and rigorous details are in von Neumann \cite{JvN39}, and 
standard operator‑algebra texts such as \cite{KR} present the GNS construction and von Neumann
  algebras in a systematic way, discussing also infinite tensor products and tail algebras.  
For physics‑oriented discussions emphasizing intuition,  examples based on infinite spin
  chains and on the thermodynamic limit are useful starting points.

\end{document}